\documentclass[conference]{IEEEtran}
\ifCLASSINFOpdf
\else
\fi
\usepackage{enumitem}
\usepackage{tikz}
\usepackage{tabularx}
\newcolumntype{Y}{>{\centering\arraybackslash}X}

\newcolumntype{Z}{>{\hsize=1.2\hsize}X}
\newcolumntype{Q}{>{\hsize=.8\hsize}X}
\newcolumntype{V}{>{\hsize=.15\hsize}X}
\usepackage{caption}
\usepackage{subcaption}
\usepackage{stfloats}
\usepackage{multirow}
\usepackage{multicol}
\usepackage{makecell}
\usepackage{balance}
\usepackage{hyperref}
\usepackage{nth}
\usepackage{xurl}
\usepackage{booktabs}

\begin{document}
%
\title{Older Adults' Experiences with Misinformation on Social Media}


\author{\IEEEauthorblockN{Filipo Sharevski}
\IEEEauthorblockA{School of Computing \\
DePaul University\\
Chicago, IL 60604\\
Email: fsharevs@depaul.edu}
\and
\IEEEauthorblockN{Jennifer Vander Loop}
\IEEEauthorblockA{School of Computing \\
DePaul University\\
Chicago, IL 60604\\
Email: jvande27@depaul.edu}
}



\maketitle

\begin{abstract}
Older adults habitually encounter misinformation on social media, but there is little knowledge about their experiences with it. In this study, we combined a qualitative survey (\textit{n}=119) with in-depth interviews (\textit{n}=21) to investigate how older adults in America conceptualize, discern, and contextualize social media misinformation. As misinformation on social media in the past was driven towards influencing voting outcomes, we were particularly interested to approach our study from a voting intention perspective. We found that 62\% of the participants intending to vote Democrat saw a manipulative political purpose behind the spread of misinformation while only 5\% of those intending to vote Republican believed misinformation has a political dissent purpose. Regardless of the voting intentions, most participants relied on source heuristics combined with fact-checking to discern truth from misinformation on social media. The biggest concern about the misinformation, among all the participants, was that it increasingly leads to biased reasoning influenced by personal values and feelings instead of reasoning based on objective evidence. The participants intending to vote Democrat were in 74\% of the cases concerned that misinformation will cause escalation of extremism in the future, while those intending to vote Republican, were undecided, or planned to abstain were concerned that misinformation will further erode the trust in democratic institutions, specifically in the context of public health and free and fair elections. During our interviews, we found that 63\% of the participants who intended to vote Republican, were fully aware and acknowledged that Republican or conservative voices often time speak misinformation, even though they are closely aligned to their political ideology.
\end{abstract}



%

\section{Introduction}
Older adults, aged 65 and up \cite{CDC2015}, are considered to be lagging behind their younger counterparts in technology and social media adoption \cite{Pang}. Despite that more than 61\% of older adults own a smartphone and only about half use social media \cite{Faverio2022}, studies show that this age group was twice more likely to be exposed to misinformation on social media than their younger counterparts during the US elections in 2016 \cite{Grinberg2019}. This age group is also generally considered less ``digitally literate,'' in that they regularly face difficulties in navigating and making use of the information they encounter online \cite{Moore2022}. Many factors are behind this digital illiteracy, be that because older adults have less knowledge of information hazards that exist online \cite{Nicholson2019}, experience usability barriers when interacting on social media such as the small text size of misinformation warnings \cite{context2022}, or have declining cognitive and physical abilities \cite{Frik2019}.

Considerable work in the intersection of older adults, technology adoption, and digital (i)literacy, so far, focused on exploring how older adults deal with security and privacy-related information online \cite{Nicholson2019, Murthy2021, Mcdonald2021}. Generally, older adults are keen to protect themselves from information and technology-related hazards, but they either lack appropriate support \cite{Morrison} or are forced to enter a collaborative enterprise to compensate for a potential loss of agency \cite{Mcdonald2021, Murthy2021}. Older adults were found to avoid using social media for information security or privacy information seeking, but they readily used social media for seeking other information \cite{Nicholson2019}

One type of information that older adults usually encounter on social media when seeking other information is \textit{misinformation}, or information lacking truth and truthfulness \cite{Jaster}. An all-encompassing term that includes various information hazards such as fabrications, rumors, conspiracies, hoaxes, fakes, clickbait, etc. \cite{Flintham}, misinformation found its way on platforms in increasingly large quantities and on a growing number of topics - from climate change, health concerns, up to elections and referendums \cite{Haughey}. Expectedly, many of the misinformation topics and much of quantities were, and perhaps still are, of interest to older adults. 

This interest together with the unpredictable consequences of social media misinformation on election results and public health, inspired small but important work in the intersection of older adults, social media adoption, and misinformation \cite{Moore2022, Brashier-Schacter, Guess2020, Wason}. Older adults, in general, might ``fall'' for fake news on social media in the short run, but they can offset this disadvantage with analytical thinking and knowledge accumulated across decades helps them evaluate claims in the long run. While this work informs the older adults' posture relative to falsehoods as information hazards, it is limited to mostly political news headlines in controlled settings. As such, it neither considers how older adults conceptualize misinformation nor how they make sense of the intentionality or the purpose of misinformation in the first place. Misinformation on social media is more than just news headlines (e.g., it might also include memes, comments, doctored videos, out-of-context posts, etc.) and past evidence suggests that peoples' mental models of misinformation differ based on what they consider to be the goal of the misinformation and who is ``behind'' it's creation and dissemination on social media \cite{folk-models}.  

To address this gap, we combined a qualitative survey (\textit{n}=119) with in-depth, semi-structured interviews (\textit{n}=21) to investigate how older adults in America conceptualize, discern, and contextualize misinformation in a broader societal context relative to their voting intentions. Past studies on misinformation extensively focused on receptivity and truth discernment relative to people's political beliefs or attitudes \cite{Pennycook1, Thorson, Sharevski-Cose}, but we chose to focus on older adults' voting intentions for the forthcoming 2024 US presidential elections instead. We made this decision for a couple of reasons. As social media misinformation in the past was driven towards influencing voting outcomes (e.g., the infamous information operations from the Internet Research Agency \cite{DiResta}), we wanted primarily to explore older adults' approach in utilizing misinformation when informing their voting intentions. We also wanted to avoid the risk of stereotyping based on people's political ideologies as recently the academic effort has been under increased political scrutiny, with claims and threats of lawsuits that the research on misinformation is driven towards suppressing conservative opinions \cite{Tollefson}.

\noindent \textbf{Results:} Our results suggest that 63\% of the participants intending to vote Democrat think that misinformation contains elements of truth for manipulative political purposes. In only 3\% of the cases, the participants with the intention to vote Republican believed misinformation has mostly a political dissent purpose, but those that intended to vote otherwise saw the purpose of misinformation to be for creating political divisions and dissent as well as manipulating emotions. Participants' voting intentions didn't factor into how they discerned truth from misinformation as they mainly relied on the source heuristic (e.g., the source of posts on social media) in combination with fact-checking. Here, we found that many participants mentioned using additional journalistic sources like the BBC to monitor the reporting on political affairs, as they believed these sources are less prone to American bias.

We found that those participants who intended to vote Democrat or weren't planning to vote were notably concerned that misinformation would escalate extremist actions and further erode the trust in formerly respected institutions. Those participants who intended to vote Republican or were undecided, on the other hand, were mostly concerned that misinformation would exacerbate the trend of motivated reasoning, i.e., see fewer people employing critical thinking. Asked about how they see the misinformation evolve in the future, most of the participants expressed skepticism that people in the US will re-establish their trust in the democratic institutions, particularly concerning the public health and free and fair elections. We also uncovered a theme where many participants, regardless of the voting intentions, were skeptical about repairing the damage to the political affairs in the US done by misinformation since the 2016 election cycle. To this point, 5 our of 8 participants who intended to vote Republican held popular Republican or conservative voices responsible of speaking misinformation frequently, regardless of the close alignment relative to their political ideology.


\noindent \textbf{Contributions}: The contributions of our work are threefold:
\begin{enumerate}
    \item Firstly, we believe this is the first work detailing the mental models older adults employ when dealing with misinformation on social media; 
    
    \item Secondly, we provide an in-depth account of how older adults experience misinformation and use this experience to inform their voting intentions;
    
    \item Finally, we report on the older adults' opinions on how misinformation has affected American society so far and how it will affect it in the future.
\end{enumerate}

\section{Background}

\subsection{Older Adults and Information Hazards Online}
Older adults' cognitive deficits, coupled with their rudimentary digital literacy and slow-to-adopt new technologies, have so far centered the research agenda mainly around how they navigate and utilize information relative to their security and privacy online \cite{Nicholson2019, Murthy2021, Mcdonald2021}. Findings suggest that this segment of the population does not trust digital sources of information to provide them with factual information when it comes to cybersecurity resources \cite{Nicholson2019}. Older adults tend to still trust broadcast media and their social resources over resources created by those with security expertise and count on collaborative family enterprises to ensure the cybersecurity of all parties \cite{Mcdonald2021}. In many cases, older adults tend to involve a ``self-appointed family tech-manager'' in their social resources who provide resources and enforcement based on their own security beliefs, causing the older adult to feel less responsible for their digital literacy \cite{Murthy2021} and less motivated to familiarize themselves with new technologies \cite{Tang2022}. 

How older adults deal with social engineering attacks (e.g., phishing, scams, etc.) has been an area of focus for much of the existing research as one of the information hazards that particularly target this part of the population. Older adults are aware of various social engineering tactics targeting the older population, especially ones thematically exploiting their critical dependence on health care services \cite{Frik2019}. But older adults tend to be overconfident in their abilities to detect phishing emails \cite{Oliveira2017} and remain vulnerable to the many evolving social engineering attacks \cite{SharevskiQRphish}. Older adults know and employ fewer phishing detection cues and frequently fail to see spam emails as unsafe \cite{Sarno}.

\subsection{Older Adults and Misinformation}
Older adults' proclivity to information hazards on social media has also inspired a small but important research agenda focused on how this population deals with fabricated, deceptive, or sensationalist news headlines. Older adults, evidence suggests, can better discern between true and fake news once they are taught skills important for verifying the credibility of online information (i.e., improving their digital literacy) and are more likely to research news stories before deciding if they were accurate or not \cite{Moore2022, Epstein, Seo2019}. Though older adults frequently use fact-checking services to correct misinformation, cognition might increasingly feel effortful for them, and they also resort to familiarity and source heuristics when evaluating news \cite{Brashier-Schacter, swire2017role} or avoid investigating misinformation altogether \cite{Geeng2020}. This part of the population usually designates one person in their social group they sought out as the expert, evidence suggests \cite{Seo2021}. 

Older adults share and interact with fake news several times more than their younger counterparts \cite{Wason, Guess2020, Grinberg2019}, but they usually stick with what they know and reject claims that contradict their knowledge even when these falsehoods feel familiar \cite{brashier2017competing}. This, coupled with their impressive knowledge, equips older adults to better distinguish fake from true headlines in the long run \cite{Pennycook1}. For older adults, it is not just important that content is misinformation, but they tend to also consider the social context, such as cues about a person’s character, more than their younger counterparts (e.g., revealing that Donald Trump averaged 15 false claims a day in 2018 may benefit older adults more than debunking any one of his `alternative facts' \cite{Brashier-Schacter}). It is believed that people on social media, when it comes to news headlines, evaluate information in a biased way to protect their political identity (i.e., motivated reasoning) and older adults are no exception to this \cite{Scherer2020}. However, this part of the population is able to offset this disadvantage because analyzing potential information hazards (e.g. misinformative headlines) increases with age \cite{hertzog2018does}.

\section{Study Design} \label{sec:methodology}
\subsection{Research Questions}
Developing an understanding of how older adults actually experience and engage with misinformation on social media is not to generalize a population, but rather to explore a phenomenon in depth. To this goal, we sought to answer the following five research questions:  

\begin{enumerate}
\itemsep 0.5em
    \item \textbf{RQ1}: How do older social media users conceptualize misinformation in general? 
    
    \item \textbf{RQ2}: How do older social media users conceptualize the purpose of misinformation?
    
    \item \textbf{RQ3}: How do older social media users discern truth from misinformation on social media?

    \item \textbf{RQ4}: How do older social media users see misinformation affecting American society in general? 

    \item \textbf{RQ5}: How do older social media users envision misinformation evolving in the future?

\end{enumerate}

\subsection{Recruitment and Sampling}
Before we started our recruitment and sampling, we obtained approval from our Institutional Review Board (IRB) to conduct an exploratory survey and follow-up semi-structured interviews (the \hyperref[sec:questionnaire]{survey questionnaire} and the \hyperref[sec:script]{interview script} are provided in the Appendix). Our sampling criteria sought participants who are social media users ages 65 and above in the United States. The criterion ``social media user'' stipulated participants to have at least one social media account and to visit at least one social media platform regularly (daily or weekly) over a period of at least a year, reading at least few posts per visit (there was no restrictions regarding posting, commenting, liking, or reporting posts on platforms as part of ``usage''). We used Prolific for the recruitment, Qualtrics for the survey, and Zoom for the follow-up interviews. At the end of the survey, we offered each of the participants to take part in a flow-up semi-structured interview. The survey was completed by a total of 119 participants (filtering out low-quality responses that didn't answered any of the questions or provided immaterial answers). Out of these 119 participants, 21 agreed to do the follow-up semi-structured interview.  

The survey responses were anonymous, but the interviews initially weren't. We offered the participants to use only the audio option in Zoom if they wanted and we recorded the meetings so we would get the automatic transcription ready for our later data analysis. Once we had the audio transcribed, we removed any personally identifiable information and checked the answers to establish full anonymity. Both the survey and the follow-up interviews allowed users to skip any question they were uncomfortable answering. The survey took around 20 minutes to complete and the follow-up interviews around 30 minutes. Participants were offered a compensation rate of \$4 for the survey and \$10 for the follow-up interview participation. The demographic structure of our survey and interview participants are given in Table \ref{tab:demographics} and Table \ref{tab:interviewdemographics}, respectively. We achieved a balanced and diverse sample, leaning towards older adults with college or graduate degrees. 

\begin{table}[tbh]
\renewcommand{\arraystretch}{1.5}
\footnotesize
\caption{Survey Demographic Distribution}
\label{tab:demographics}
\centering
\begin{tabularx}{\linewidth}{|Y|}
\hline
 \textbf{Gender} \\\hline
\footnotesize
\vspace{0.2em}
    \hfill \makecell{\textbf{Female} \\ 64 (54\%)} 
    \hfill \makecell{\textbf{Male} \\ 54 (45\%)} 
    \hfill \makecell{\textbf{Non-cisgender} \\ 1 (1\%)} \hfill\null
\vspace{0.2em}
\\\hline
 \textbf{Intent to Vote} \\\hline
\footnotesize
\vspace{0.2em}
    \hfill \makecell{\textbf{Democrat} \\ 68 (57\%)} 
    \hfill \makecell{\textbf{Republican} \\ 29 (25\%)} 
    \hfill \makecell{\textbf{Undecided} \\ 17 (14\%)} 
    \hfill \makecell{\textbf{No Intention to Vote} \\ 5 (4\%)} \hfill\null 
\vspace{0.2em}
\\\hline
\textbf{Race/Ethnicity} \\\hline
\footnotesize
\vspace{0.2em}
     \hfill \makecell{\textbf{Asian} \\ 1 (1\%)} 
     \hfill \makecell{\textbf{Black/African American}\\ 17 (14\%)} 
     \hfill \makecell{\textbf{Latinx} \\ 2 (2\%)} 
    \hfill \makecell{\textbf{White} \\ 96 (81\%)}
     \hfill \makecell{\textbf{More than One} \\ 3 (2\%)} 
 \vspace{0.2em}
 \\\hline
 \textbf{Level of Education} \\\hline
\footnotesize
\vspace{0.2em}
    \hfill \makecell{\textbf{High school} \\ 12 (10\%)} 
    \hfill \makecell{\textbf{College} \\ 74 (62\%)} 
    \hfill \makecell{\textbf{Graduate} \\ 33 (28\%)} \hfill\null 
\vspace{0.2em}
\\\hline
\end{tabularx}
\end{table}

\begin{table}[tbh]
\renewcommand{\arraystretch}{1.5}
\footnotesize
\caption{Interview Demographic Distribution}
\label{tab:interviewdemographics}
\centering
\begin{tabularx}{\linewidth}{|Y|}
\hline
 \textbf{Gender} \\\hline
\footnotesize
\vspace{0.2em}
    \hfill \makecell{\textbf{Female} \\ 11 (52\%)} 
    \hfill \makecell{\textbf{Male} \\ 9 (43\%)} 
    \hfill \makecell{\textbf{Non-cisgender} \\ 1 (5\%)} \hfill\null
\vspace{0.2em}
\\\hline
 \textbf{Intent to Vote} \\\hline
\footnotesize
\vspace{0.2em}
    \hfill \makecell{\textbf{Democrat} \\ 12 (57\%)} 
    \hfill \makecell{\textbf{Republican} \\ 8 (38\%)} 
    \hfill \makecell{\textbf{Undecided} \\ 1 (5\%)} 
\hfill\null 
\vspace{0.2em}
\\\hline
\textbf{Race/Ethnicity} \\\hline
\footnotesize
\vspace{0.2em}
     \hfill \makecell{\textbf{Black/African American}\\ 2 (10\%)} 
     \hfill \makecell{\textbf{Latinx} \\ 1 (5\%)} 
    \hfill \makecell{\textbf{White} \\ 17 (80\%)}
     \hfill \makecell{\textbf{More than One} \\ 1 (5\%)} 
 \vspace{0.2em}
 \\\hline
 \textbf{Level of Education} \\\hline
\footnotesize
\vspace{0.2em} 
    \hfill \makecell{\textbf{College} \\ 13 (62\%)} 
    \hfill \makecell{\textbf{Graduate} \\ 8 (38\%)} \hfill\null 
\vspace{0.2em}
\\\hline
\end{tabularx}
\end{table}

\subsection{Data Collection}
Past studies have extensively analyzed the interplay between misinformation and people's political beliefs or attitudes \cite{folk-models}. Recently, this academic effort has come under increased political scrutiny, with claims and threats of lawsuits that the research on misinformation is driven toward suppressing conservative opinions \cite{Tollefson}. To avoid the risk of further stereotyping relative to people's self-reported political stances, we chose to collect and group our participants by their \textit{intention to vote} in the forthcoming 2024 presidential election cycle. This way, we can allow for a better analysis of the political contextualization of misinformation without risks of misinterpretation as well as allow for situations where people who identify with one political ideology intend to vote for another party, remain undecided, or abstain.

Much of the prior research has also focused on older adults within specific regions by contacting older adults through senior centers \cite{Frik2019, Seo2021}. While this is an efficient method to gather data on older adults, we felt that a more widespread study would provide additional information on older adults who have at least some form of digital literacy and are more likely to be habitual social media users. In addition, we felt it was important to include a larger, more diverse population of older adults because this aspect is often overlooked in other studies focused on older adults \cite{Hargittai2017}. Capturing a diverse part of the population allowed us to provide an ecological validity to our qualitative findings relative to the older adults' misinformation mental models (concept and purpose), truth discernment strategies, and misinformation outlook.

\subsection{Data Analysis}
Our data analysis process involved several steps. First, we collated the interview responses with the responses from the survey, assigned numbers to the participants, and made a final check for the anonymity of the overall answers. Next, we performed an inductive coding approach \cite{Saldana.2013, Thomas.2006} to identify frequent, dominant, or significant aspects in the collated answers of our participants. One of the researchers open-coded all the data. The resulting codes were then discussed with a second coder, who then independently coded all the data on their own. We calculated the \textit{Inter-Rater Reliability} (IRR) indicator using Cohen's kappa \cite{Cohen.1960} to determine the level of agreement of the two coders, reaching a $k = 0.90$, which we deemed acceptable. 

The open codes were then used to structure a codebook that captures four main aspects: (i) \textit{conceptualization} i.e., codes pertaining to how older users mentally model misinformation in general; (ii) \textit{purpose} i.e., codes related to the purposes of misinformation that older adults believe are behind its spread on social media;  (iii) \textit{truth discernment} i.e., codes describing the participants' cognitive strategies for discerning truth in the information they encounter on social media; and (iv) \textit{misinformation effects} i.e., what older adults believe are the effects of misinformation on the American society in general. The \hyperref[sec:codebook]{overall codebook} is listed in the Appendix. 
\section{Results}
\subsection{RQ1: Misinformation Mental Models}
Past research shows that people's mental models of misinformation differ relative to whether they see misinformation as mainly falsehoods and inaccuracies that could easily be verified or fact-checked or a more nuanced form of deception that might contain elements of true and accurate information \cite{folk-models}. This distinction is important because in the latter case, the deception could be seen as a deliberate attempt to exercise influencing power over how people form opinions or decide about actions \cite{Chadwick}. For this research question, we stopped short of explicitly asking about the intent behind the deception as we wanted to get more detailed and rich responses with the second research question. Therefore, here we only captured the way our participants conceptualized misinformation respective to a) \textit{falsehoods and inaccuracies} (e.g., entirely lacking truthfulness); and b) \textit{misleading or deceiving claims} (e.g., including elements of truth) \cite{Jaster}.

\begin{figure}[!h]
  \centering
\centering\includegraphics[width=\linewidth]{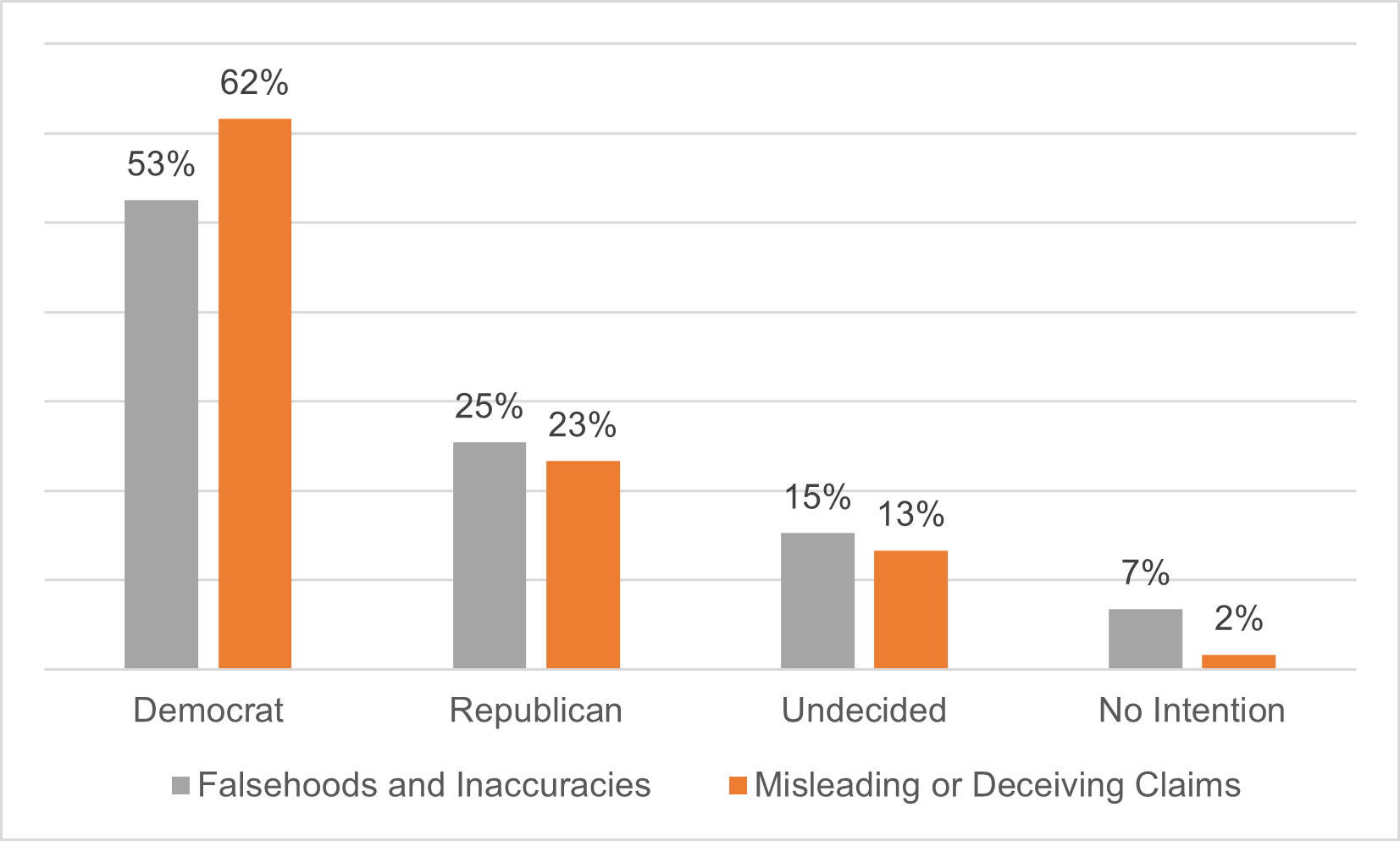}
\caption{\textbf{RQ1}: Mental Models vs. Intention to Vote}
  \label{fig:rq1}
\end{figure}

We broke down the coded answers from both our survey and our follow-up interviews controlling for the participants' intention to vote in Figure \ref{fig:rq1}. Though our sample is dominated by participants intending to vote Democrat in total, they make up 62\% of the overall participants who think that misinformation contains elements of truth. For example, participant \textbf{P111}  offered the following concept of misinformation during the follow-up interview: 

\begin{quote}
    ``\textit{Misinformation is `information' that is either entirely or partially untrue. It can be presented in a manner that makes it difficult to discern its veracity. Misinformation can be deliberately shared to convert people to a particular way of thinking, or it can be information taken out of context. Most often bias is involved. In the past, reputable publishers conducted fact-checking to limit the amount of misinformation shared, but the demise of traditional journalism and the spread of social media gives everyone a soapbox. People lacking the credentials to self-edit or those who deliberately seek to put out false info to promote their own agendas are quite prevalent.}''
\end{quote}

For most of the participants with voting intentions other than Democrat, misinformation was mainly falsehoods or inaccuracies. For example, participant \textbf{P40} simply stated that for them, misinformation is ``\textit{any information that has been verified not to be true}.'' It is worth mentioning that we noticed a trend among the participants who intended to vote Republican to speak about ``\textit{disinformation}.'' Here, \textbf{P35} offered this detailed account of disinformation: 

\begin{quote}
``\textit{The word disinformation brings to my mind a government or reporting organization sponsoring or composing facts. Because I can remember back in the Cold War we, being the Western countries, would come out with probably disinformation about what was happening in Russia. And certainly, from what I can gather, the Russian media was the same in the other direction. They're very happy to use, I hesitate to use the word propaganda, but I think that propaganda and disinformation intersect in that Venn diagram of reporting. To me, disinformation is a deliberate act by somebody. It takes a lot of planning to fabricate information and there is an agenda of some sort behind it.}''    
\end{quote}

\subsection{RQ2: Purpose of Misinformation}
While in the first research question we were interested to learn whether older adults see elements of truth or not in misinformation on social media, in the second research question we wanted our participants to explicitly unpack the purpose of misinformation. Whether content is disseminated deliberately or not, studies differentiate between \textit{misinformation} (no intentionality) and \textit{disinformation} (intentionality). But determining intentionality is often difficult \cite{swire2020public}, so we thus considered a more encompassing aspect of misinformation, i.e. its \textit{purpose} on social media. This gave us an opportunity to ask our participants to elaborate more not just about whether there is intentionality or not, but who actually creates, disseminates, and benefits from misinformation in their opinions. 

Past research offers an abundance of evidence to demonstrate that the purpose of misinformation is to create divisions and dissent along political lines \cite{Stewart, Zannettou}, perpetuate political counter-argumentation \cite{folk-models}, and ultimately undermine political candidates for office \cite{DiResta}. But misinformation was also used and continues to be used outside of the political arena such as in health misinformation \cite{vanderLinden, Ghenai}, consumer product misinformation \cite{Zeng}, and alternative narratives in general (e.g., conspiracies, rumors) \cite{Chandra, Geeng}. Equally, misinformation is an essential ingredient for a general propagandistic rhetoric of emotionally manipulative language \cite{Huffaker}. Therefore, we analyzed the purpose of misinformation respective to a) \textit{creation of divisions and/or dissent} (i.e., politically-related); b) \textit{influence opinion and/or action} (e.g., mainly for personal or financial gain); and c) \textit{manipulate emotions} (i.e., mainly for manipulating emotions). 

\begin{figure}[!h]
  \centering
\centering\includegraphics[width=\linewidth]{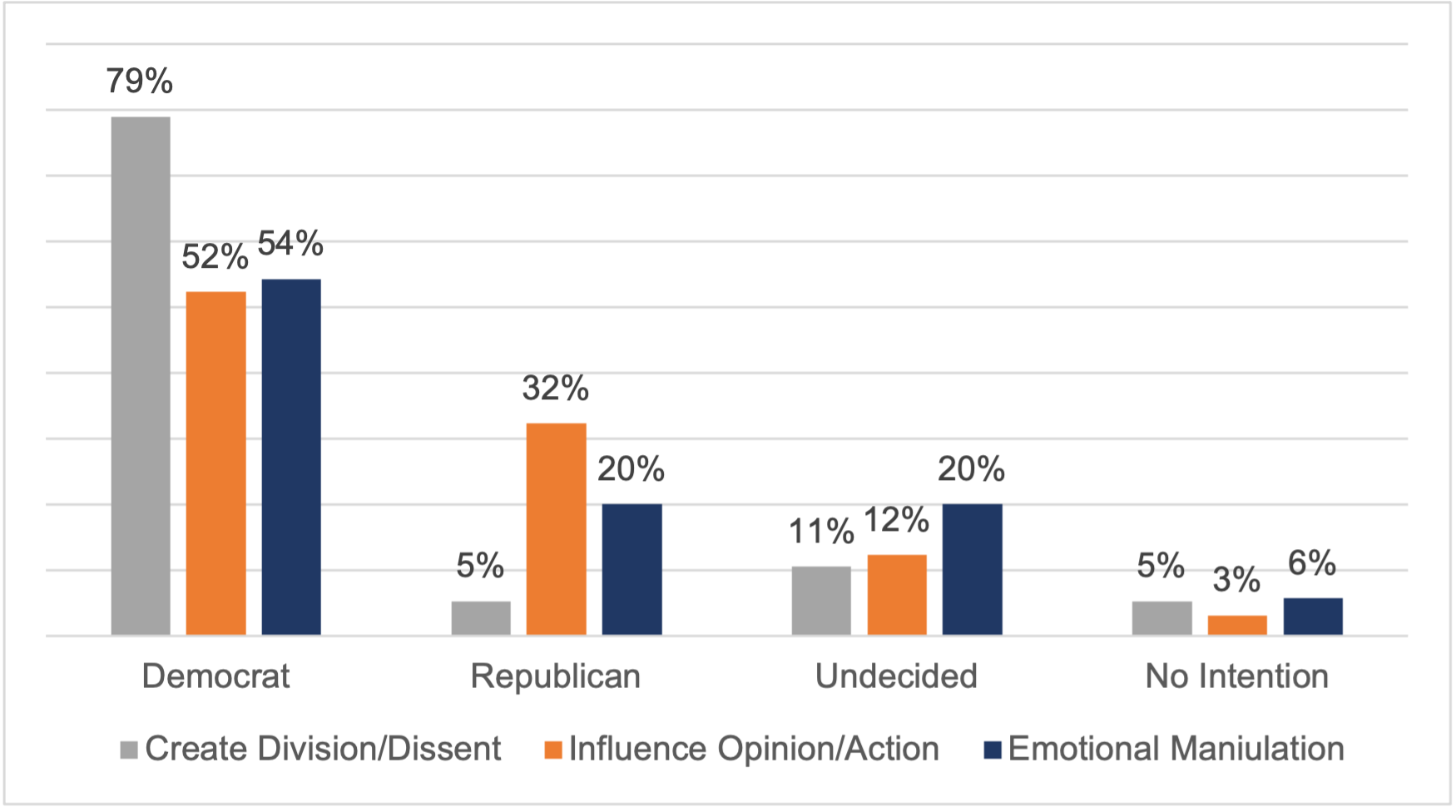}
\caption{\textbf{RQ2}: Purpose of Misinformation s vs. Intention to Vote}
  \label{fig:rq2}
\end{figure}

As shown in Figure \ref{fig:rq2}, regardless of the intent to vote, the majority of each group mostly saw that misinformation served the purpose of influencing opinions and actions. For example, \textbf{P100}, who intended to vote Democrat, acknowledged the political aspect of misinformation but offered a broader view on influencing opinions and/or actions: 

\begin{quote}
    ``\textit{Yes, I think the two primary purposes are 1) to profit financially from misinformation that encourages others to purchase products, and 2) to ``educate'' others about issues the spreader thinks are important. This would include those who want to swing others to adopting specific views (e.g., on vaccines or abortion), voting for specific candidates, etc.}''
\end{quote}

However, we noticed that the participants with the intention to vote Republican in most of the cases concentrated on influencing opinions and actions or manipulating emotions as the main purpose of misinformation compared to only 5\% who believed misinformation has mostly a divisive/dissent purpose. For example, \textbf{P47} offered this take on the intent behind the spread of misinformation on social media: 

\begin{quote}
    ``\textit{Of course the spread is purposeful. People distribute misinformation to hopefully change your mind about something or they want to gain power, money or whatever it is they desire to change. If you think social media is bad, try reading or watching any news articles that are totally biased.}''
\end{quote}

We also noticed that the participants with the intention to vote Democrat form the vast majority of those participants who believe misinformation's main purpose is creating political divisions and/or dissent with 79\%. Participant \textbf{P63} believed that ``\textit{political groups intend to divide people by spreading misinformation towards influencing election results, fomenting racism, and inciting violence}.'' Another participant, \textbf{P36}, stated that the purpose of misinformation is ``\textit{to sway popular opinion, to cause dissent, to widen rifts, to gain votes, to stir up trouble}. One of the participants we interviewed, \textbf{P10}, offered a bit more of a detailed take:

\begin{quote}
``\textit{Well, in the case of Russia, it's been documented that they had not just a casual intent to spread misinformation on social media. I mean, there was a real campaign to get Donald Trump elected in 2016, and you know, probably also in 2020. But I know in 2016 it was definitely an orchestrated campaign. Yeah, I think that there's always intent behind misinformation. I mean, they're not just doing it for kicks. I think that you know they're trying to mislead someone about an outlandish political candidate.}''
\end{quote}

Participant \textbf{P82} traced back the antecedents of the circumstances in which misinformation materializes its divisive and dissenting purpose: 

\begin{quote}
    ``\textit{A lot of misinformation is designed by the corporations, the politicians, the billionaires to really just distract us and have us fighting among ourselves so they can just go and do whatever they want, make more money, control policy. And behind this is the intention of dumbing down us Americans. It started back in the sixties when the Nixon Administration purposely started under funding public education. The kids that were raised in the sixties in the public school system, you know how old are they now? They're my age. And we are the last generation that knows what critical thinking is. After us, the public schools taught an entire population to question nothing. And that's why misinformation can get to these people because they're like, well, that sounds good. No, it doesn't. No, it doesn't.}''
\end{quote}

Relative to the emotional manipulation purpose of misinformation, we noticed that the participants who were undecided were more prevalent with 20\% despite their smaller share in the overall sample. One of them, \textbf{P17}, stated that misinformation is ``\textit{designed to gravitate towards higher emotions and become more adamant about feeling that person's theories about everything are always valid regardless if they actually are or aren't}.'' Those with no intention to vote were more leaning towards the emotional manipulation  purpose with participant \textbf{P57} noting that ``\textit{in the US the purpose is to mislead and confuse, to leave the populace in ignorance so they can be exploited and made complacent}''.

We further queried and interviewed them to ask about who they think creates and disseminates misinformation and who benefits from it. The participants who were uncertain or who had no intention to vote pointed mainly to individual social media ``\textit{influencers}'' as they either ``\textit{want credit for posting it},'' according to participant \textbf{P2} or ``\textit{gain notoriety from the shock ``value'' and the sheer malicious enjoyment of it},'' according to participant \textbf{P37}. The participants who intended to vote Democrat mainly blamed the ``other side'' of the political divide, or ``\textit{Trump and his supporters, white supremacists, Fox, right-wing media (Bannon, Jones), anti-gun control, Republican governors (Texas, Florida, etc.).}'' according to participant \textbf{P18}. Participant \textbf{P26}, interestingly, offered a bit more balanced and extended answer: 

\begin{quote}
    ``\textit{I see it most among conservative politicians and their followers. Trump followers are habitual misinformation hawkers but there are liberals on my side of the aisle that practice it too. QAnon followers are another group that heavily traffics in misinformation}''
\end{quote}

We noticed that the participants who intended to vote Republican were keen to blame the social media companies as the creators and distributors of misinformation. Participant \textbf{P108} pointed out ``\textit{the `higher ups' in social media companies because they refuse to censor or correct information while deliberately permitting misinformation to be disseminated to attack their opponents}''. Participant \textbf{P24} extended this viewpoint as: 

\begin{quote}
``\textit{The majority has been done by left-leaning organizations trying to keep the truth from different events and situations from coming out and of course they are helped by the algorithms and people who work hard to censor those who have a different viewpoint. The exception is now X.}''    
\end{quote}

As to who benefits from the misinformation on social media, those who intended to vote Democrat mostly pointed to ``\textit{big business and the GOP}'', as participant \textbf{P15} put it. Participant \textbf{P38} extended this list to ``\textit{foreign governments trying to weaken the US; Political demagogic political operatives; Corporate interests like Exxon and Phillip Morris who flood the media with false information about the grave dangers of fossil fuels on climate or tobacco on public health, respectively.}'' Those participants that intended to vote Republican, were uncertain, or planned not to vote saw the ``\textit{government}'' as the main beneficiary of misinformation in addition to ``\textit{big tech and big pharma}, '' according to participant \textbf{P101}. In the words of participant \textbf{P44}, the ``\textit{governmental agencies and the corporate elite benefit from misinformation because it keeps the public in the dark or they provide wrong data to get the public to do what they want}.''





\subsection{RQ3: Truth Discernment}
With our third research question we wanted to learn how older adults discern between misinformation and truthful statements on social media. Past research suggests that poor truth discernment is linked to a lack of analytical reasoning and relevant knowledge, as well as to the use of familiarity and source heuristics \cite{Pennycook-Rand-Psych}. At some point, it was believed that people are ``better'' at discerning truth from falsehood (despite greater overall belief) when evaluating politically concordant news \cite{VanBavel, Kahan}. However, more and more evidence showed that ``better'' truth discernment is associated with analytical reasoning, regardless of whether a statement is consistent or inconsistent with one's partisanship \cite{Bago, Bronstein}. To this end, we probed our participants about the cues and the reasoning they employ when assessing content on social media or whether they mainly employed a) \textit{heuristics} (e.g., familiarity or source); b) \textit{analytical thinking} (e.g., they reflected or deliberated about the content); and c) \textit{a mix of both strategies}.

\begin{figure}[!h]
  \centering
\centering\includegraphics[width=\linewidth]{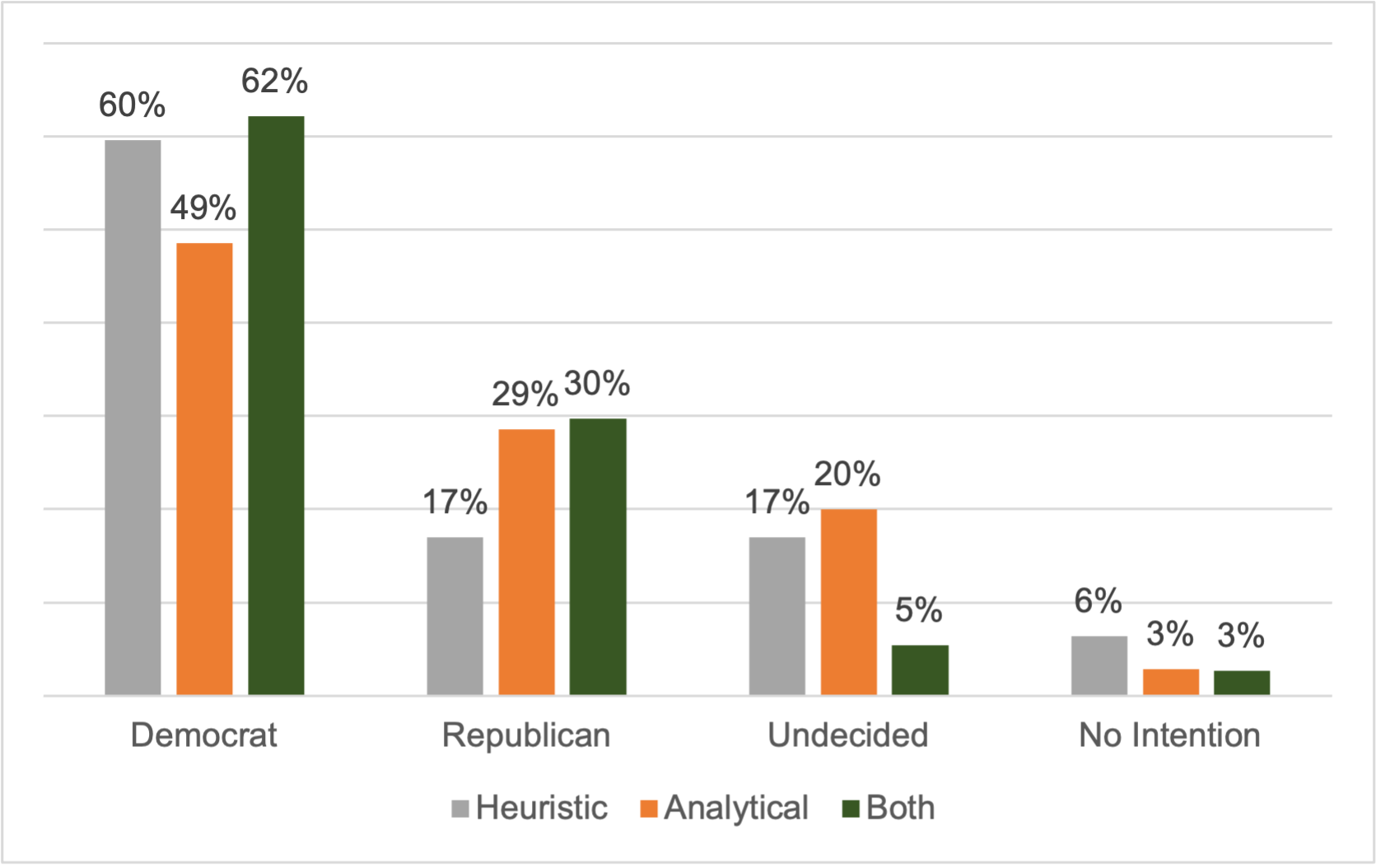}
\caption{\textbf{RQ3}: Truth Discernment vs. Intention to Vote}
  \label{fig:rq3}
\end{figure}

The main theme that emerged from both our survey and the interviews was that regardless of the intention to vote -- shown in Figure \ref{fig:rq3} -- our participants mainly relied on the source heuristic in combination with fact-checking to discern truth from misinformation. Interestingly, many participants mentioned they use additional sources like the BBC to also check the reporting on US politics because, as participant \textbf{P39} put it, ``\textit{they are less prone to American bias}''. The main rule of thumb, perhaps, was the use of familiarity with the topic and the source heuristics to look for cues of fabrication or deception. Participant \textbf{P26} who intended to vote Democrat explicitly referred to the source heuristics:   

\begin{quote}
    ``\textit{I look at the source of the information in terms of who is stating the information. I know if it originates from a political source not to take it at face value. Same for partisan influencers. Even comments from private individuals make me look back at their other comments to see what they've said in the recent past. Doesn't take long to see who is ultra-partisan and who is not.}'' 
\end{quote}

Another participant from this group, \textbf{P100}, elaborated on their combined tactics for discernment between misinformation and truthful statements based on their familiarity with emotionally manipulative language: 

\begin{quote}
``\textit{I look for emotion-based, sweeping statements that offer no factual evidence or that offer evidence I don't see as credible. Especially if they're trying to sell something or stir up enmity toward certain groups, races, public health workers, public officials, etc. I'm alert for suggestions of violence against others or property. If I'm not sure about a claim, I check with sources I'm certain are credible, such as the CDC, state or local public health departments, evidence-based science journals, my city councilor, health care professionals, etc.}''    
\end{quote}

Participant \textbf{P62} explicitly mentioned the use of fact-checking services as part of their varied analytical approach in spotting any content that might be entirely lacking or omitting good parts of the truth:

\begin{quote}
    ``\textit{I try to use mainstream media such as the Washington Post, or the New York Times for information that I want to be factual. I also keep up with the local news and carefully choose which affiliate I am watching. It did not take me long to figure out that broadcast owners broadcast what is in their best interest and TV can be slanted to the liberal or conservative side. I recently signed up for a fact-checking newsletter to stay better informed about what is or is not the truth. I think it is very easy to become misinformed by social media simply because so many people with so many different points of view are participating with each one having their own agenda.}''
\end{quote}

The participants who intended to vote Republican employed the same strategies for truth discernment as their Democratic counterparts. Participant \textbf{P35} offered an interesting approach using the source heuristics: 

\begin{quote}
    ``\textit{Confirmation bias is always a problem here, where information that generally accords with one's own beliefs is likely to be accepted even if it includes glaring inaccuracies. I look to see if claims are patently ridiculous, i.e., Donald Trump eats babies. The spelling and grammar of a printed source are as important as sloppy work may conceal sloppy facts. What is the source?  Are they generally reliable?  Are they positioned to the left or right of the political spectrum? Finally, does it coincide with what I reasonably believe to be true?}''
\end{quote}

A noticeable portion of the participants who intended to vote Republican indicated they regularly turn to fact-checking services. Using their familiarity with emotionally manipulative language in combination with reflective reasoning, participant \textbf{P117} said that ``\textit{fact-checking sites are a great way to check the credibility of news I come across because stories that have a lot of emotional manipulation tend to be false}.'' Another participant, \textbf{P85} seconded this approach, adding: 

\begin{quote}
    ``\textit{I look to see who posted the misinformation - do they post only about one thing or how many followers do they have?  If something does not sound right, I will do a fact check on Snopes or Politifact}'' 
\end{quote} 

The participants who were undecided or intended not to vote were also keen on using the source heuristics or turning to fact-checking services, many of them  invoking the lack of critical thinking in sharing fake news on social media. Participant \textbf{P37} pointed out that they know a post is misinformation ``\textit{when it has a high level of stridency, it has absolutely nothing to do with critical or factual thinking, and people persistently post the same content (verbatim) from other ``sources''}. Another participant, \textbf{P54}, amended this strategy, stating:

\begin{quote}
    ``\textit{I always look for credible sources, or I will use Politico or Snopes or others to vet it myself. I think it is vital nowadays to be very discerning when reading about something on social media. Look at who posts the info, is it from a credible source, and even if it is someone you trust, a quick search to vet it is almost necessary to do so in these times.}''
\end{quote}

\subsection{RQ4: Misinformation and Society}
\begin{figure}[!h]
  \centering
\centering\includegraphics[width=\linewidth]{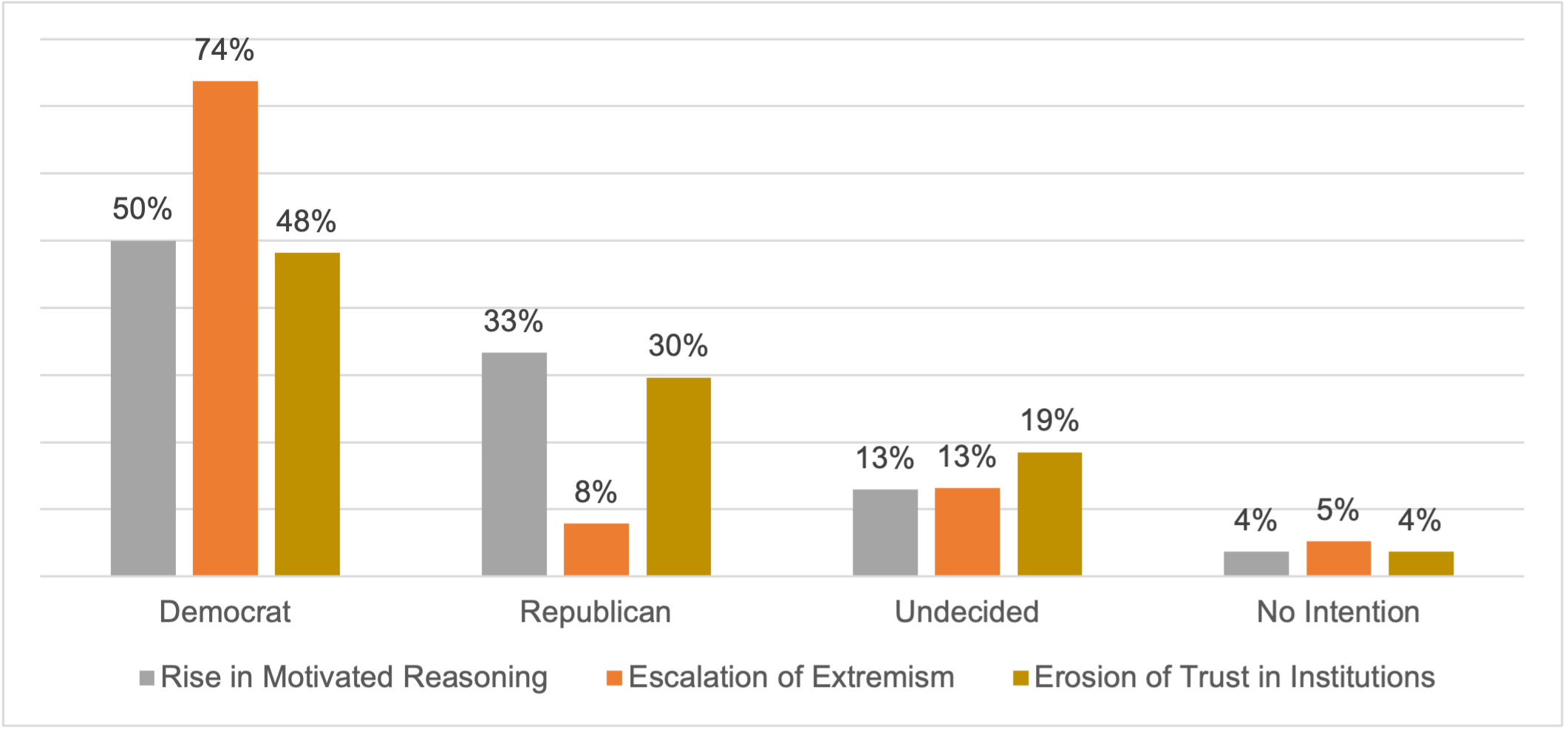}
\caption{\textbf{RQ4}: Misinformation and Society vs. Intention to Vote}
  \label{fig:rq4}
\end{figure}

With the fourth research question, we probed our participants on their opinions on how misinformation, or the general trend of truth decay, affects American society in general. The truth decay phenomenon is associated with a set of several specific trends, namely: (a) \textit{motivated reasoning} or reasoning driven more by personal or moral values than objective evidence and analytical interpretations of data \cite{Kahan}; (b) \textit{escalation of extremism} - or the collaborative construction and amplification of alleged evidence of general foul play (e.g., election fraud) used to facilitate extremist action \cite{Prochaska}; and (c) \textit{trust erosion} or diminishing trust in formerly respected institutions as sources of factual information \cite{rich2018truth}. Respective to these specific trends, we summarized the coded answers of our participants' combined answers in Figure \ref{fig:rq4}. 

As to the intention to vote, we noticed that those participants who intended to vote Democrat were in 74\% on the opinion that misinformation will escalate extremist actions, but less concerned that it will exacerbate the trend of motivated reasoning (50\%) and further and erode the trust in formerly respected institutions (48\%). Participant \textbf{P19} expressed the following concerns: 

\begin{quote}
    ``\textit{It's dangerous to society. During the pandemic, it probably caused illness and more deaths by misleading people as to how they needed to protect themselves against COVID-19. In politics it can cause the rise of extremist leaders and governments, convince people of false election results, suggest that people having done despicable things they didn't, or cause people to think someone is innocent who is a criminal and dangerous.}''
\end{quote}

Another participant, \textbf{P36} expressed concerns about the erosion of trust in institutions in the context of widening the divisions among Americans:  

\begin{quote}
    ``\textit{I think misinformation divides our society more and more, widening the gap in political differences. It also makes people not trust the government, the media, education, science, and more. It causes us to fight against each other, rather than join forces for the common good.}''
\end{quote}

Those participants who intended to vote Republican or were undecided were mostly concerned that misinformation will both exacerbate the trend of motivated reasoning as well as further erode the trust in the institutions. Participant \textbf{P35} offered the following opinion: 

\begin{quote}
    ``\textit{I think society is much worse off because of disinformation. Before the Internet, if you wanted to fact-check something you went to the library and looked it up.  Since more effort was involved in writing a book, and publishers valued their reputations, there was an incentive to get things right.  Now anyone can write anything and instantly have an audience of potentially millions.  Some percentage of that audience will be stupid and blithely pass along any old trash that appeals to them.  Society as a whole is composed of people from all walks of life and if a certain number of them believe things that are just wrong it is bound to have a negative effect on any collective decisions in the long run.}''
\end{quote}

\subsection{RQ5: Misinformation in the Future}
We came up with our final research question after we analyzed the results of the survey and identified that older adults as a target demographic could offer rich reflections on how misinformation might evolve in the future, based on their unique experience with misinformation -- in any form -- over a longer period of time and even preceding the Internet and social media. Therefore, we decided to ask those participants who opted for a follow-up interview to expand upon how misinformation might evolve relative to a) potential \textit{public health crises in the future}; and b) \textit{future election cycles} starting with the forthcoming one in 2024. 

The main theme we noticed among our participants was skepticism that people in the US will re-establish their trust in the institutions, particularly the health authorities. As one of the demographics that suffered the most during the COVID-19 pandemic, our participants used their experience with the virus and vaccination misinformation \cite{swire2020public, Sharevski-Cose} to reckon about future public health misinformation. In these regards, participant \textbf{P35} stated:  

\begin{quote}
    ``\textit{I think that because of what happened during COVID people are much more skeptical now about what they are being told to do and not do. I guess more specifically if the government says, run down and get a vaccination or always wear a mask or never, ever, ever leave your house or try not to breathe. I believe there's going to be a lot more skepticism involved, and I think if it's another COVID which is going to be with us forever, everybody will catch it eventually. Everybody will build up some sort of immunity to it, but I think if it's something far more serious, like Marburg Virus Disease or some of the viruses that are emerging in North Africa and are now popping up around the world, there won't be the public response to effectively quarantine, vaccinate and defend against it because of the mistrust that a lot of people now feel.}''
\end{quote}





Participant \textbf{P14}, invoking the politicization of the COVID-19 pandemic \cite{Hart} and the specific trend of rejection of science \cite{rich2018truth}, pointed to the antecedents of the general health misinformation as a setup that will enable it to continue negatively affecting public health with a very honest account: 

\begin{quote}
    ``\textit{[Misinformation] initially increased a lot of disbelief in science and will continue to do so. I think one of the ways I look at it is public health is not the strong side of the United States because you have so many institutions against each other. It was the CDC against the NIH, and this one here, and this one there. I am an immunologist and of course I believe in vaccinations, but I was skeptical when the government said it would take a year to develop a COVID-19 vaccine because previously it would take 5 or 6 years to develop one. Early on in the vaccine development and testing process, I thought all of it might well be misinformation, but the mistake was that neither the CDC nor the NIH were open to further simple explanations and transparency to let people know that it was a development and learning in progress and that the vaccine will anyhow evolve over a longer period time. So, these misinformation mistakes might happen again in the future and cause another disaster.}''
\end{quote}

Focused on the forthcoming 2024 election and evoking the past two election cycles, most of the participants could not avoid involving Donald Trump when they surmised about the election misinformation future. For these participants, he may have had (or will have) a temporary role in the misinformation conundrum but he has done permanent damage to overall political affairs in the US, especially with his brand of motivated reasoning. Participant \textbf{P23}, to this end, stated: 

\begin{quote}
    ``\textit{I'm frightened. I think misinformation will very possibly incite violence. There's not going to be any truth coming from anywhere, because the truth is just off the table. Now, nobody cares. People want to feel like they know what's going on, but they don't. I think it's very frustrating for everyone. If someone tells you, they know what's going on politically, automatically you know that you probably can discount whatever they say. It's just not true. We live in a greed-based society that kind of drives everyone to use misinformation to get what they want. And it will get worse in every future election cycle.}''
\end{quote}

\section{Discussion}
Our findings indicate older adults are fairly well attuned to the current state of misinformation on social media. Although we didn't explicitly test any particular misinformation content nor employ any truth discernment interventions, we found additional evidence that older adults do extensively research both news stories as well as other types of content before deciding if they were accurate or not \cite{Moore2022, Epstein, Seo2019}. We noticed a strong determination to use fact-checking sites and, interestingly, foreign journalistic services in combination with source and familiarity heuristics \cite{Brashier-Schacter, swire2017role}. Though we had a sample with a majority of older adults with college or graduate degrees, none of them indicated they avoid investigating misinformation altogether \cite{Geeng2020}, but instead offered a detailed account of their truth discernment strategies, which heavily relied on independent analytical thinking (i.e., they sought experts \textit{outside} of their social groups, not \textit{inside}, as the prior evidence suggests \cite{Seo2021}). To these points, participant \textbf{P57} noted: 

\begin{quote}
``\textit{Yes. I typically trust Snopes.com to be unbiased and fair. There are others, like Factcheck.com. Otherwise, I fact-check myself by visiting multiple well-known news sites, and try to read things from central, left-leaning, and right-leaning sources to keep a fair balance.}''
\end{quote}

Perhaps anecdotal, but telling nonetheless, we found evidence that older adults reject claims that contradict their knowledge even when these falsehoods feel familiar, similarly to what previously was reported for older adults relative to misinformation \cite{brashier2017competing}. Equally, older adults, as the past evidence suggest \cite{Brashier-Schacter}, did also consider the social context, such as cues about a person’s character, our findings show. During our follow-up interviews, 5 out of 8 participants who intended to vote Republican, stated they are fully aware that Republican or conservative voices often time speak misinformation, even though they are closely aligned to their political ideology. Participant \textbf{P93} honestly remarked:

\begin{quote}
``\textit{I like Trump. I thought he did a good job, but you know he's misleading people. He is telling them, `They are after you'. There's nobody after you or me. So that's misinformation. I don't even think he's smart enough to even concoct that up. I think somebody wrote that for him and said, `say this and your poll numbers will come up'. Okay, that's misinformation.}''   
\end{quote}



We can't say for sure that the older adults in our sample didn't resort to evaluating the information on social media in a biased way to protect their political identity \cite{Scherer2020} nor if they are better at distinguishing fake from true headlines in the long run \cite{Pennycook1, hertzog2018does}. But, concordant with the evidence in \cite{folk-models}, older adults do, in considerable numbers, model misinformation as a form of political (counter)argumentation that often contains elements of truth. Along the lines of the evidence in \cite{Lima} and \cite{folk-models}, the older adults in our sample indeed saw the ``other side'' of the political divide as responsible for misinformation on social media, even though they acknowledged that there are spreaders ``\textit{on their side of the aisle}'', as participant \textbf{P26} said. The concerns about the more pronounced tie between extreme polarization and misinformation adherence found in \cite{Efstratiou} were also expressed in our study. Altogether, the outlook offered by the participants in our study seems to be rather non-optimistic as many of them were skeptical of restoring the democratic order they remember from the past.

\subsection{Ethical Considerations}
The purpose of our study was not to generalize to a population; rather, to explore how older adults, who are social media users, experience misinformation in their daily lives. While we added the participants' self-reported gender, race/ethnicity, and level of education for a transparent reporting of our recruitment and sampling procedures, we avoided providing definitive numbers accompanying results with these demographics. Instead, we supported our findings with descriptive quotations from participants that convey the way older adult social media users experience misinformation in the hope that the results will make a meaningful contribution to the study of misinformation on social media as a whole. To minimize any risk of misinterpretation or adverse effects on the participants' intentions to vote, at the end of the survey and the interview we performed an extensive debriefing to point out that we the researchers are politically impartial, we are not affiliated with any social media platform in any way, and that we are not advocating for any misinformation action such as content moderation or account removal. 

Ethical concerns do arise when dealing with misinformation, as the probability of harmful implication is non-negligible within a pluralistic social media population including older adults. We didn't exposed our participants to misleading statements in the survey, but references to misleading content did come up during our semi-structured interviews related to COVID-19 and past election cycles. We were clear and open to directly point out to our participants that we cannot comment on any of these topics in order to eliminate a risk of conceiving or perpetuating any misconceptions about these actively debated topics on social media. We also asked our participants before we started with the interview questions that we don't require any social media content to be accessed to avoid 
a knock-on effect on their future experiences on these platforms (i.e., a situation where platforms might begin recommend more questionable content or use this access event to re-tailor any target advertising). To ensure we obtained correct understanding of our participants' experiences, we reviewed the main points we recorded during the interview and clarified any misunderstandings we might have. We also sent a draft of our paper to our participants for feedback.

\subsection{Limitations}
Our research was limited in its scope to US social media users ages 65 and above. A limitation also comes from the sampling method, the use of a survey provider, and a mix of survey and follow-up interviews. Other qualitative studies with a different sample than ours might yield variations in the mental models of misinformation, differences in conceptualizations of the misinformation's intentionality, as well as different strategies of truth discernment. Our sample, though balanced on the other demographic variables, nonetheless mostly consisted of older adults with college or graduate degrees. Older adults with other levels or education might have a very different experience with misinformation on social media and we acknowledge that this is a limitation to our study. We also didn't test any of the self-reported mental models and truth discernment strategies with actual social media content. This limits the results from generalization to current or future content where older adults might well readjust or change their truth discernment strategies. 

Our study is also limited to the state of misinformation on social media during the first half of 2023 as well as the platforms that currently older adults use. Future platforms, or changes in the policies of platforms, might affect how older adults perceive and discern misinformation from the truth (e.g., many right-leaning users saw Twitter as perpetuating left-leaning ``misinformation'' and infringing on the freedom of speech in the past \cite{gettr-paper}, but participants in our study noted they see that ``\textit{since transforming to X, the `other side' is usually able to post information rebutting this misinformation},'' per participant \textbf{P44}, whose intention was to vote Republican)). Though we left our participants sufficient time to express their opinions, answer their questions, and comment on aspects they deemed important,  nonetheless, the time allowed might have not been insufficient for them to formulate a more informed expression about their overall experience with misinformation on social media. Despite all these limitations, and similar to mixed and qualitative studies in general, we believe our results provide rich accounts of older adults' lived experiences with misinformation on a scale and comprehensiveness that have not been done before.

\subsection{Future Work}
We plan to continue our work on how older adults experience and interact with misinformation, expanding our sampling to account for older adults' with other levels of education. During our interviews, participants discussed moderation efforts that they have encountered while using various social media platforms. As these moderation efforts were \cite{Saltz, gettr-paper}, and still are \cite{Thorson}, subject of political contention, we plan not just to explore their posture relative to their voting intentions, but also consider novel platforms such as TikTok where older adults increasingly participate and encounter misinformation \cite{sharevski2023abortion}. Older adults often use assistive technologies and we are equally interested in getting further understanding on how accessibility shapes their experiences with misinformation. We are equally interested in exploring other perspectives than intention to vote of information seeking online in general and social media in particular beyond. To this point, we also plan to work with older adults in improving and testing ''misinformation literacy'' interventions towards a better truth discernment not just for this particular population but for anyone on social media in general. 

\section{Conclusion}
Misinformation undoubtedly shapes the way older adults make decisions in their daily lives, including for whom they intend to vote or whether to follow particular medical advice. Though this part of the population might lack the cognitive abilities and digital literacy to deal with the abundance of misinformation content across the entire social media landscape, our results suggest that they are fairly well-attuned and prepared to deal with it. We found that the intentions to vote of our participants factored in how they made sense of misinformation on social media, but they were largely irrelevant to how they discerned truth from misinformation in general. Misinformation is undoubtedly an evolving concept that, in the view of the older adults in our study, is a serious threat to the democratic institutions in the US and a problem that might take a long time to rectify.



%

\bibliographystyle{IEEEtran}
\bibliography{senior-disinfo}

\begin{thebibliography}{10}
\providecommand{\url}[1]{#1}
\csname url@samestyle\endcsname
\providecommand{\newblock}{\relax}
\providecommand{\bibinfo}[2]{#2}
\providecommand{\BIBentrySTDinterwordspacing}{\spaceskip=0pt\relax}
\providecommand{\BIBentryALTinterwordstretchfactor}{4}
\providecommand{\BIBentryALTinterwordspacing}{\spaceskip=\fontdimen2\font plus
\BIBentryALTinterwordstretchfactor\fontdimen3\font minus \fontdimen4\font\relax}
\providecommand{\BIBforeignlanguage}[2]{{%
\expandafter\ifx\csname l@#1\endcsname\relax
\typeout{** WARNING: IEEEtran.bst: No hyphenation pattern has been}%
\typeout{** loaded for the language `#1'. Using the pattern for}%
\typeout{** the default language instead.}%
\else
\language=\csname l@#1\endcsname
\fi
#2}}
\providecommand{\BIBdecl}{\relax}
\BIBdecl

\bibitem{CDC2015}
\BIBentryALTinterwordspacing
N.~C. for Chronic Disease~Prevention and H.~Promotion, ``Indicator definitions - older adults,'' 2015. [Online]. Available: \url{https://www.cdc.gov/cdi/definitions/older-adults.html#:~:text=Category%3A%20Older%20Adults-,Demographic%20Group%3A,persons%20aged%20%E2%89%A565%20years}
\BIBentrySTDinterwordspacing

\bibitem{Pang}
\BIBentryALTinterwordspacing
C.~Pang, Z.~Collin~Wang, J.~McGrenere, R.~Leung, J.~Dai, and K.~Moffatt, ``Technology adoption and learning preferences for older adults: Evolving perceptions, ongoing challenges, and emerging design opportunities,'' in \emph{Proceedings of the 2021 CHI Conference on Human Factors in Computing Systems}, ser. CHI '21.\hskip 1em plus 0.5em minus 0.4em\relax New York, NY, USA: Association for Computing Machinery, 2021. [Online]. Available: \url{https://doi.org/10.1145/3411764.3445702}
\BIBentrySTDinterwordspacing

\bibitem{Faverio2022}
\BIBentryALTinterwordspacing
M.~Faverio, ``Share of those 65 and older who are tech users has grown in the past decade,'' Jan 2022. [Online]. Available: \url{https://www.pewresearch.org/short-reads/2022/01/13/share-of-those-65-and-older-who-are-tech-users-has-grown-in-the-past-decade/}
\BIBentrySTDinterwordspacing

\bibitem{Grinberg2019}
N.~Grinberg, K.~Joseph, L.~Friedland, B.~Swire-Thompson, and D.~Lazer, ``Fake news on twitter during the 2016 u.s. presidential election,'' \emph{Science}, vol. 363, no. 6425, pp. 374--378, 2019.

\bibitem{Moore2022}
R.~C. Moore and J.~T. Hancock, ``A digital media literacy intervention for older adults improves resilience to fake news,'' \emph{Scientific Reports}, vol.~12, no.~1, p. 6008, 2022.

\bibitem{Nicholson2019}
\BIBentryALTinterwordspacing
J.~Nicholson, L.~Coventry, and P.~Briggs, ``"if it's important it will be a headline": Cybersecurity information seeking in older adults,'' in \emph{Proceedings of the 2019 CHI Conference on Human Factors in Computing Systems}, ser. CHI '19.\hskip 1em plus 0.5em minus 0.4em\relax New York, NY, USA: Association for Computing Machinery, 2019, p. 1–11. [Online]. Available: \url{https://doi-org.ezproxy.depaul.edu/10.1145/3290605.3300579}
\BIBentrySTDinterwordspacing

\bibitem{context2022}
F.~Sharevski, A.~Devine, P.~Jachim, and E.~Pieroni, ``Meaningful context, a red flag, or both? preferences for enhanced misinformation warnings among us twitter users,'' in \emph{Proceedings of the 2022 European Symposium on Usable Security}, ser. EuroUSEC '22.\hskip 1em plus 0.5em minus 0.4em\relax New York, NY, USA: Association for Computing Machinery, 2022, p. 189–201, \url{https://doi.org/10.1145/3549015.3555671}.

\bibitem{Frik2019}
\BIBentryALTinterwordspacing
A.~Frik, L.~Nurgalieva, J.~Bernd, J.~Lee, F.~Schaub, and S.~Egelman, ``Privacy and security threat models and mitigation strategies of older adults,'' in \emph{Fifteenth Symposium on Usable Privacy and Security (SOUPS 2019)}.\hskip 1em plus 0.5em minus 0.4em\relax Santa Clara, CA: USENIX Association, Aug. 2019, pp. 21--40. [Online]. Available: \url{https://www.usenix.org/conference/soups2019/presentation/frik}
\BIBentrySTDinterwordspacing

\bibitem{Murthy2021}
\BIBentryALTinterwordspacing
S.~Murthy, K.~S. Bhat, S.~Das, and N.~Kumar, ``Individually vulnerable, collectively safe: The security and privacy practices of households with older adults,'' \emph{Proc. ACM Hum.-Comput. Interact.}, vol.~5, no. CSCW1, apr 2021. [Online]. Available: \url{https://doi.org/10.1145/3449212}
\BIBentrySTDinterwordspacing

\bibitem{Mcdonald2021}
\BIBentryALTinterwordspacing
N.~Mcdonald and H.~M. Mentis, ``“citizens too”: Safety setting collaboration among older adults with memory concerns,'' \emph{ACM Trans. Comput.-Hum. Interact.}, vol.~28, no.~5, aug 2021. [Online]. Available: \url{https://doi.org/10.1145/3465217}
\BIBentrySTDinterwordspacing

\bibitem{Morrison}
\BIBentryALTinterwordspacing
B.~Morrison, L.~Coventry, and P.~Briggs, ``How do older adults feel about engaging with cyber-security?'' \emph{Human Behavior and Emerging Technologies}, vol.~3, no.~5, pp. 1033--1049, 2023/09/02 2021. [Online]. Available: \url{https://doi.org/10.1002/hbe2.291}
\BIBentrySTDinterwordspacing

\bibitem{Jaster}
R.~Jaster and D.~Lanius, ``Speaking of fake news,'' \emph{The epistemology of fake news}, vol.~19, 2021.

\bibitem{Flintham}
\BIBentryALTinterwordspacing
M.~Flintham, C.~Karner, K.~Bachour, H.~Creswick, N.~Gupta, and S.~Moran, ``Falling for fake news: Investigating the consumption of news via social media,'' in \emph{Proceedings of the 2018 CHI Conference on Human Factors in Computing Systems}, ser. CHI '18.\hskip 1em plus 0.5em minus 0.4em\relax New York, NY, USA: Association for Computing Machinery, 2018, p. 1–10. [Online]. Available: \url{https://doi.org/10.1145/3173574.3173950}
\BIBentrySTDinterwordspacing

\bibitem{Haughey}
\BIBentryALTinterwordspacing
M.~McClure~Haughey, M.~Povolo, and K.~Starbird, ``Bridging contextual and methodological gaps on the “misinformation beat”: Insights from journalist-researcher collaborations at speed,'' in \emph{Proceedings of the 2022 CHI Conference on Human Factors in Computing Systems}, ser. CHI '22.\hskip 1em plus 0.5em minus 0.4em\relax New York, NY, USA: Association for Computing Machinery, 2022. [Online]. Available: \url{https://doi.org/10.1145/3491102.3517503}
\BIBentrySTDinterwordspacing

\bibitem{Brashier-Schacter}
N.~M. Brashier and D.~L. Schacter, ``Aging in an era of fake news,'' \emph{Current Directions in Psychological Science}, vol.~29, no.~3, pp. 316--323, 2020.

\bibitem{Guess2020}
A.~M. Guess, M.~Lerner, B.~Lyons, J.~M. Montgomery, B.~Nyhan, J.~Reifler, and N.~Sircar, ``A digital media literacy intervention increases discernment between mainstream and false news in the united states and india,'' \emph{Proceedings of the National Academy of Sciences}, vol. 117, no.~27, pp. 15\,536--15\,545, 2020.

\bibitem{Wason}
\BIBentryALTinterwordspacing
M.~Wason, S.~S. Gupta, S.~Venkatraman, and P.~Kumaraguru, ``Building sociality through sharing: Seniors' perspectives on misinformation,'' in \emph{Proceedings of the 10th ACM Conference on Web Science}, ser. WebSci '19.\hskip 1em plus 0.5em minus 0.4em\relax New York, NY, USA: Association for Computing Machinery, 2019, p. 321–322. [Online]. Available: \url{https://doi.org/10.1145/3292522.3326052}
\BIBentrySTDinterwordspacing

\bibitem{folk-models}
F.~Sharevski, A.~Devine, E.~Pieroni, and P.~Jachim, ``{Folk Models of Misinformation On Social Media},'' in \emph{{Network and distributed system security symposium}}, 2023, \url{https://dx.doi.org/10.14722/ndss.2023.24293}.

\bibitem{Pennycook1}
\BIBentryALTinterwordspacing
G.~Pennycook and D.~G. Rand, ``Lazy, not biased: Susceptibility to partisan fake news is better explained by lack of reasoning than by motivated reasoning,'' \emph{Cognition}, vol. 188, pp. 39--50, 2019, the Cognitive Science of Political Thought. [Online]. Available: \url{https://www.sciencedirect.com/science/article/pii/S001002771830163X}
\BIBentrySTDinterwordspacing

\bibitem{Thorson}
E.~Thorson, ``Belief echoes: The persistent effects of corrected misinformation,'' \emph{Political Communication}, vol.~33, no.~3, pp. 460--480, 2016.

\bibitem{Sharevski-Cose}
F.~Sharevski, R.~Alsaadi, P.~Jachim, and E.~Pieroni, ``Misinformation warnings: Twitter's soft moderation effects on covid-19 vaccine belief echoes,'' \emph{Computers \& Security}, vol. 114, p. 102577, 2022.

\bibitem{DiResta}
R.~DiResta, K.~Shaffer, B.~Ruppel, D.~Sullivan, R.~Matney, R.~Fox, J.~Albright, and B.~Johnson, ``The tactics \& tropes of the internet research agency,'' 2019.

\bibitem{Tollefson}
\BIBentryALTinterwordspacing
{Tollefson, Jeff}, ``Disinformation researchers under investigation: what’s happening and why,'' 2023. [Online]. Available: \url{https://www.nature.com/articles/d41586-023-02195-3}
\BIBentrySTDinterwordspacing

\bibitem{Tang2022}
\BIBentryALTinterwordspacing
X.~Tang, Y.~Sun, B.~Zhang, Z.~Liu, R.~LC, Z.~Lu, and X.~Tong, ``"i never imagined grandma could do so well with technology": Evolving roles of younger family members in older adults' technology learning and use,'' \emph{Proc. ACM Hum.-Comput. Interact.}, vol.~6, no. CSCW2, nov 2022. [Online]. Available: \url{https://doi.org/10.1145/3555579}
\BIBentrySTDinterwordspacing

\bibitem{Oliveira2017}
\BIBentryALTinterwordspacing
D.~Oliveira, H.~Rocha, H.~Yang, D.~Ellis, S.~Dommaraju, M.~Muradoglu, D.~Weir, A.~Soliman, T.~Lin, and N.~Ebner, ``Dissecting spear phishing emails for older vs young adults: On the interplay of weapons of influence and life domains in predicting susceptibility to phishing,'' in \emph{Proceedings of the 2017 CHI Conference on Human Factors in Computing Systems}, ser. CHI '17.\hskip 1em plus 0.5em minus 0.4em\relax New York, NY, USA: Association for Computing Machinery, 2017, p. 6412–6424. [Online]. Available: \url{https://doi.org/10.1145/3025453.3025831}
\BIBentrySTDinterwordspacing

\bibitem{SharevskiQRphish}
F.~Sharevski, A.~Devine, E.~Pieroni, and P.~Jachim, ``{Phishing with Malicious QR Codes},'' in \emph{European Symposium on Usable Security}, ser. EuroUSEC `22.\hskip 1em plus 0.5em minus 0.4em\relax New York, NY, US: ACM, 2022, pp. 160--171.

\bibitem{Sarno}
D.~M. Sarno, J.~E. Lewis, C.~J. Bohil, and M.~B. Neider, ``Which phish is on the hook? phishing vulnerability for older versus younger adults,'' \emph{Human Factors}, vol.~62, no.~5, pp. 704--717, 2023/09/04 2019.

\bibitem{Epstein}
\BIBentryALTinterwordspacing
Z.~Epstein, G.~Pennycook, and D.~Rand, ``Will the crowd game the algorithm? using layperson judgments to combat misinformation on social media by downranking distrusted sources,'' in \emph{Proceedings of the 2020 CHI Conference on Human Factors in Computing Systems}, ser. CHI '20.\hskip 1em plus 0.5em minus 0.4em\relax New York, NY, USA: Association for Computing Machinery, 2020, p. 1–11. [Online]. Available: \url{https://doi.org/10.1145/3313831.3376232}
\BIBentrySTDinterwordspacing

\bibitem{Seo2019}
\BIBentryALTinterwordspacing
H.~Seo, J.~Erba, D.~Altschwager, and M.~Geana, ``Evidence-based digital literacy class for older, low-income african-american adults,'' \emph{Journal of Applied Communication Research}, vol.~47, no.~2, pp. 130--152, 2019. [Online]. Available: \url{https://doi.org/10.1080/00909882.2019.1587176}
\BIBentrySTDinterwordspacing

\bibitem{swire2017role}
B.~Swire, U.~K. Ecker, and S.~Lewandowsky, ``The role of familiarity in correcting inaccurate information.'' \emph{Journal of experimental psychology: learning, memory, and cognition}, vol.~43, no.~12, p. 1948, 2017.

\bibitem{Geeng2020}
\BIBentryALTinterwordspacing
C.~Geeng, S.~Yee, and F.~Roesner, ``Fake news on facebook and twitter: Investigating how people (don't) investigate,'' in \emph{Proceedings of the 2020 CHI Conference on Human Factors in Computing Systems}, ser. CHI '20.\hskip 1em plus 0.5em minus 0.4em\relax New York, NY, USA: Association for Computing Machinery, 2020, p. 1–14. [Online]. Available: \url{https://doi.org/10.1145/3313831.3376784}
\BIBentrySTDinterwordspacing

\bibitem{Seo2021}
H.~Seo, M.~Blomberg, D.~Altschwager, and H.~T. Vu, ``Vulnerable populations and misinformation: A mixed-methods approach to underserved older adults' online information assessment,'' \emph{New Media \& Society}, vol.~23, no.~7, pp. 2012--2033, 2021.

\bibitem{brashier2017competing}
N.~M. Brashier, S.~Umanath, R.~Cabeza, and E.~J. Marsh, ``Competing cues: Older adults rely on knowledge in the face of fluency.'' \emph{Psychology and aging}, vol.~32, no.~4, p. 331, 2017.

\bibitem{Scherer2020}
\BIBentryALTinterwordspacing
P.~Scherer, Laura~D. and P.~Pennycook, Gordon, ``\BIBforeignlanguage{English}{Who is susceptible to online health misinformation?}'' \emph{\BIBforeignlanguage{English}{American Journal of Public Health, suppl.Supplement 3}}, vol. 110, pp. S276--S277, 10 2020, name - University of Colorado; Copyright - Copyright American Public Health Association Oct 2020; Last updated - 2023-03-01; SubjectsTermNotLitGenreText - United States--US; India. [Online]. Available: \url{https://ezproxy.depaul.edu/login?url=https://www.proquest.com/scholarly-journals/who-is-susceptible-online-health-misinformation/docview/2531706561/se-2}
\BIBentrySTDinterwordspacing

\bibitem{hertzog2018does}
C.~Hertzog, R.~M. Smith, and R.~Ariel, ``Does the cognitive reflection test actually capture heuristic versus analytic reasoning styles in older adults?'' \emph{Experimental aging research}, vol.~44, no.~1, pp. 18--34, 2018.

\bibitem{Hargittai2017}
\BIBentryALTinterwordspacing
E.~Hargittai and K.~Dobransky, ``Old dogs, new clicks: Digital inequality in skills and uses among older adults,'' \emph{Canadian Journal of Communication}, vol.~42, p. 195–212, 2017. [Online]. Available: \url{https://doi.org/10.22230/cjc.2017v42n2a3176}
\BIBentrySTDinterwordspacing

\bibitem{Saldana.2013}
J.~Salda\~na, \emph{The coding manual for qualitative researchers}.\hskip 1em plus 0.5em minus 0.4em\relax Los Angeles, CA, US: SAGE, 2013.

\bibitem{Thomas.2006}
D.~R. Thomas, ``A {General} {Inductive} {Approach} for {Analyzing} {Qualitative} {Evaluation} {Data},'' \emph{American Journal of Evaluation}, vol.~27, no.~2, pp. 237--246, 2006.

\bibitem{Cohen.1960}
J.~Cohen, ``A coefficient of agreement for nominal scales,'' \emph{Educational and Psychological Measurement}, vol.~20, no.~1, pp. 37--46, 1960.

\bibitem{Chadwick}
\BIBentryALTinterwordspacing
A.~Chadwick and J.~Stanyer, ``{Deception as a Bridging Concept in the Study of Disinformation, Misinformation, and Misperceptions: Toward a Holistic Framework},'' \emph{Communication Theory}, vol.~32, no.~1, pp. 1--24, 10 2021. [Online]. Available: \url{https://doi.org/10.1093/ct/qtab019}
\BIBentrySTDinterwordspacing

\bibitem{swire2020public}
B.~Swire-Thompson, D.~Lazer \emph{et~al.}, ``Public health and online misinformation: challenges and recommendations,'' \emph{Annu Rev Public Health}, vol.~41, no.~1, pp. 433--451, 2020.

\bibitem{Stewart}
L.~G. Stewart, A.~Arif, and K.~Starbird, ``Examining trolls and polarization with a retweet network,'' in \emph{Proc. ACM WSDM, workshop on misinformation and misbehavior mining on the web. 2018.}, 2018.

\bibitem{Zannettou}
\BIBentryALTinterwordspacing
S.~Zannettou, T.~Caulfield, W.~Setzer, M.~Sirivianos, G.~Stringhini, and J.~Blackburn, ``Who let the trolls out? towards understanding state-sponsored trolls,'' in \emph{Proceedings of the 10th ACM Conference on Web Science}, ser. WebSci ’19.\hskip 1em plus 0.5em minus 0.4em\relax New York, NY, USA: Association for Computing Machinery, 2019, p. 353–362. [Online]. Available: \url{https://doi.org/10.1145/3292522.3326016}
\BIBentrySTDinterwordspacing

\bibitem{vanderLinden}
S.~van~der Linden, ``Misinformation: susceptibility, spread, and interventions to immunize the public,'' \emph{Nature Medicine}, vol.~28, no.~3, pp. 460--467, 2022.

\bibitem{Ghenai}
\BIBentryALTinterwordspacing
A.~Ghenai and Y.~Mejova, ``Fake cures: User-centric modeling of health misinformation in social media,'' \emph{Proc. ACM Hum.-Comput. Interact.}, vol.~2, no. CSCW, nov 2018. [Online]. Available: \url{https://doi.org/10.1145/3274327}
\BIBentrySTDinterwordspacing

\bibitem{Zeng}
\BIBentryALTinterwordspacing
E.~Zeng, T.~Kohno, and F.~Roesner, ``What makes a “bad” ad? user perceptions of problematic online advertising,'' in \emph{Proceedings of the 2021 CHI Conference on Human Factors in Computing Systems}, ser. CHI '21.\hskip 1em plus 0.5em minus 0.4em\relax New York, NY, USA: Association for Computing Machinery, 2021. [Online]. Available: \url{https://doi.org/10.1145/3411764.3445459}
\BIBentrySTDinterwordspacing

\bibitem{Chandra}
\BIBentryALTinterwordspacing
P.~Chandra and J.~Pal, ``Rumors and collective sensemaking: Managing ambiguity in an informal marketplace,'' in \emph{Proceedings of the 2019 CHI Conference on Human Factors in Computing Systems}, ser. CHI '19.\hskip 1em plus 0.5em minus 0.4em\relax New York, NY, USA: Association for Computing Machinery, 2019, p. 1–12. [Online]. Available: \url{https://doi.org/10.1145/3290605.3300563}
\BIBentrySTDinterwordspacing

\bibitem{Geeng}
\BIBentryALTinterwordspacing
C.~Geeng, S.~Yee, and F.~Roesner, ``Fake news on facebook and twitter: Investigating how people (don't) investigate,'' in \emph{Proceedings of the 2020 CHI Conference on Human Factors in Computing Systems}, ser. CHI '20.\hskip 1em plus 0.5em minus 0.4em\relax New York, NY, USA: Association for Computing Machinery, 2020, p. 1–14. [Online]. Available: \url{https://doi.org/10.1145/3313831.3376784}
\BIBentrySTDinterwordspacing

\bibitem{Huffaker}
\BIBentryALTinterwordspacing
J.~S. Huffaker, J.~K. Kummerfeld, W.~S. Lasecki, and M.~S. Ackerman, ``Crowdsourced detection of emotionally manipulative language,'' in \emph{Proceedings of the 2020 CHI Conference on Human Factors in Computing Systems}, ser. CHI '20.\hskip 1em plus 0.5em minus 0.4em\relax New York, NY, USA: Association for Computing Machinery, 2020, p. 1–14. [Online]. Available: \url{https://doi.org/10.1145/3313831.3376375}
\BIBentrySTDinterwordspacing

\bibitem{Pennycook-Rand-Psych}
G.~Pennycook and D.~G. Rand, ``The psychology of fake news,'' \emph{Trends in Cognitive Sciences}, vol.~25, no.~5, pp. 388--402, 2021.

\bibitem{VanBavel}
J.~J. Van~Bavel and A.~Pereira, ``The partisan brain: An identity-based model of political belief,'' \emph{Trends in cognitive sciences}, vol.~22, no.~3, pp. 213--224, 2018.

\bibitem{Kahan}
D.~M. Kahan, ``Misconceptions, misinformation, and the logic of identity-protective cognition,'' 2017.

\bibitem{Bago}
B.~Bago, D.~G. Rand, and G.~Pennycook, ``Fake news, fast and slow: Deliberation reduces belief in false (but not true) news headlines.'' \emph{Journal of experimental psychology: general}, 2020.

\bibitem{Bronstein}
M.~V. Bronstein, G.~Pennycook, A.~Bear, D.~G. Rand, and T.~D. Cannon, ``Belief in fake news is associated with delusionality, dogmatism, religious fundamentalism, and reduced analytic thinking,'' \emph{Journal of Applied Research in Memory and Cognition}, vol.~8, no.~1, pp. 108--117, 2019.

\bibitem{Prochaska}
\BIBentryALTinterwordspacing
S.~Prochaska, K.~Duskin, Z.~Kharazian, C.~Minow, S.~Blucker, S.~Venuto, J.~D. West, and K.~Starbird, ``Mobilizing manufactured reality: How participatory disinformation shaped deep stories to catalyze action during the 2020 u.s. presidential election,'' \emph{Proc. ACM Hum.-Comput. Interact.}, vol.~7, no. CSCW1, apr 2023. [Online]. Available: \url{https://doi.org/10.1145/3579616}
\BIBentrySTDinterwordspacing

\bibitem{rich2018truth}
M.~D. Rich \emph{et~al.}, \emph{Truth decay: An initial exploration of the diminishing role of facts and analysis in American public life}.\hskip 1em plus 0.5em minus 0.4em\relax Rand Corporation, 2018.

\bibitem{Hart}
P.~S. Hart, S.~Chinn, and S.~Soroka, ``Politicization and polarization in covid-19 news coverage,'' \emph{Science Communication}, vol.~42, no.~5, pp. 679--697, 2020.

\bibitem{Lima}
\BIBentryALTinterwordspacing
G.~Lima, J.~Han, and M.~Cha, ``Others are to blame: Whom people consider responsible for online misinformation,'' \emph{Proc. ACM Hum.-Comput. Interact.}, vol.~6, no. CSCW1, apr 2022. [Online]. Available: \url{https://doi.org/10.1145/3512953}
\BIBentrySTDinterwordspacing

\bibitem{Efstratiou}
\BIBentryALTinterwordspacing
A.~Efstratiou and E.~De~Cristofaro, ``Adherence to misinformation on social media through socio-cognitive and group-based processes,'' \emph{Proc. ACM Hum.-Comput. Interact.}, vol.~6, no. CSCW2, nov 2022. [Online]. Available: \url{https://doi.org/10.1145/3555589}
\BIBentrySTDinterwordspacing

\bibitem{gettr-paper}
F.~Sharevski, A.~Devine, P.~Jachim, and E.~Pieroni, ``{``Gettr-ing'' User Insights from the Social Network Gettr},'' 2022, \url{https://truthandtrustonline.com/wp-content/uploads/2022/10/TTO\_2022_proceedings.pdf}.

\bibitem{Saltz}
\BIBentryALTinterwordspacing
E.~Saltz, C.~R. Leibowicz, and C.~Wardle, ``Encounters with visual misinformation and labels across platforms: An interview and diary study to inform ecosystem approaches to misinformation interventions,'' in \emph{Extended Abstracts of the 2021 CHI Conference on Human Factors in Computing Systems}, ser. CHI EA '21.\hskip 1em plus 0.5em minus 0.4em\relax New York, NY, USA: Association for Computing Machinery, 2021. [Online]. Available: \url{https://doi.org/10.1145/3411763.3451807}
\BIBentrySTDinterwordspacing

\bibitem{sharevski2023abortion}
F.~Sharevski, J.~V. Loop, P.~Jachim, A.~Devine, and E.~Pieroni, ``Abortion misinformation on tiktok: Rampant content, lax moderation, and vivid user experiences,'' \emph{arXiv preprint arXiv:2301.05128}, 2023.

\end{thebibliography}

\section{Study Questionnaire} \label{sec:questionnaire}
\subsection*{Conceptualization and Purpose \textbf{[Open Ended]}} 

\begin{enumerate}

        \item Can you please \textbf{define} ``misinformation'' in your own words? Please be elaborate in details. 
        
        \item Who \textbf{creates} misinformation on social media, in your opinion? Please be elaborate in details. 
        
        \item Who \textbf{disseminates} misinformation on social media, in your opinion? Please be elaborate in details. 
        
        \item Who \textbf{benefits} from misinformation on social media, in your opinion? Please be elaborate in details. 
        
        \item Do you think there is a \textbf{purpose} behind the spread of misinformation on social media, in your opinion? Please be elaborate in details. 

        \item What \textbf{cues} do you use to determine if a social media post is a misinformation? Please be elaborate in details. 
        
        
        
        


        \item How does misinformation \textbf{affect the society} in general, in your opinion? Please be elaborate in details. 
\end{enumerate}

\section{Interview Script} \label{sec:script}
\begin{enumerate}

\item We asked you how do you define misinformation. Could you please elaborate more about the way you yourself conceptualize it or make sense of it?

\item We asked you about the intent behind the spread of misinformation. Could you please elaborate more about the broader context of how you see this intent?

\item What is your take on misinformation during the COVID-19 Pandemic?

\item In your opinion, how misinformation might evolve and affect future health crises? 

\item What is your take on misinformation during the election cycles 2016 and 2020? 

\item In your opinion, how misinformation will evolve and affect the forthcoming US 2024 elections? 

\end{enumerate}

\section{Codebook} \label{sec:codebook}
\begin{itemize}
    \item \textbf{Conceptualization} Codes pertaining to the how older users mentally model misinformation in general:
    \begin{itemize}

            \item \textbf{Falsehoods and Inaccuracies} The participant defined misinformation as information that lacks truth and truthfulness \cite{Jaster}
            \item \textbf{Misleading or Deceiving Statement} The participant defined misinformation as information that contains elements of truth as to mislead or deceive \cite{folk-models}
           
    \end{itemize}

    \item \textbf{Purpose} Codes pertaining to how older adults see the purposes of misinformation behind its spread on social media:
    \begin{itemize}

            \item \textbf{Create Divisions / Dissent} The participant expressed that misinformation's main purpose is to create divisions and cause political dissent \cite{Stewart, Zannettou}
         
            \item \textbf{Influence Opinions and Actions} The participant expressed that misinformation's main purpose is to influence opinions and actions \cite{vanderLinden}

            \item \textbf{Propagandize} The participant expressed that misinformation's main purpose is to support propagandistic efforts \cite{Huffaker}
           
    \end{itemize}

    \item \textbf{Truth Discernment} Codes pertaining to how older adults discern  truth from misinformation on social media
    \begin{itemize}

            \item \textbf{Heuristics} The participant expressed that they mainly employ heuristics, e.g., cognitive shortcuts, intuition, and feelings \cite{Pennycook-Rand-Psych}
         
            \item \textbf{Analytical Thinking} The participant expressed that they mainly employ analytical thinking \cite{Bago}

            \item \textbf{Both} The participant expressed that they employ a mix of heuristics and analytical thinking 
           
    \end{itemize}

    \item \textbf{Misinformation Effects} Codes pertaining to the older adults' perspectives on the effects of misinformation on the American society in general
    \begin{itemize}

            \item \textbf{Rise in Motivated Reasoning} The participant expressed that they see negative effects such as a rise in motivated reasoning across the entire population of the US \cite{Kahan}
         
            \item \textbf{Escalation of Extremism} The participant expressed that they see negative effects such as escalation of extremism within the US \cite{Pang}

            \item \textbf{Erosion of Trust in Institutions} The participant expressed that they see negative effects such as continued erosion in trust of the governing institutions in the US \cite{Pennycook1}
           
    \end{itemize}

\end{itemize}

\end{document}